\documentclass[aps,prl,showpacs,showkeys,twocolumn]{revtex4}
\usepackage{mathrsfs}
\usepackage{amssymb}
\usepackage{amsmath}
\usepackage{bm}
\usepackage{graphics}
\usepackage{natbib}

\begin{document}
\title{Suppressed Josephson phase transition in one parallel double-quantum-dot junction}
\author{Guang-Yu Yi$^1$}
\author{Xiao-Qi Wang$^1$}
\author{Cui Jiang$^2$}
\author{Wei-Jiang Gong$^1$}\email{gwj@mail.neu.edu.cn}

\affiliation{1. College of Sciences, Northeastern University, Shenyang
110819, China\\
2. Basic Department, Shenyang Institute of Engineering, Shenyang 110136, China}

\date{\today}

\begin{abstract}
With the help of the numerical renormalization group method, we theoretically investigate the Josephson phase transition in a parallel junction with one quantum dot embedded in each arm. It is found that in the cases of uniform dot levels and dot-superconductor couplings, the Josephson phase transition will be suppressed. This is manifested as the fact that with the enhancement of the electron correlation, the supercurrent only arrives at its $\pi'$ phase but cannot enter its $\pi$ phase. Moreover, when the dot levels are detuned, one $\pi'$-phase island appears in the phase diagram. Such a result is attributed to the nonlocal motion of the Cooper pair in this structure. We believe that this work can be helpful in understanding the Josephson phase transition modified by the electron correlation and quantum interference.
\end{abstract}
 \maketitle

\bigskip

\emph{Introduction}---
The successful fabrication of quantum dots (QDs) allows scientists to investigate the conventional electron correlation in the mesoscopic circuits, due to their strong Coulomb repulsion and shiftable levels. It has been found that the Kondo effect, the most typical electron correlation, indeed take effect to the electron tunneling\cite{kondo1,kondo2,kondo3}. Moreover, when one QD is introduced in one Josephson junction, the strong electron interaction drives the well-known Josephson phase transition\cite{jpt1,jpt2,jpt3,jpt4}. If the Kondo temperature $T_K$ is larger than the superconducting pairing energy  $\Delta$, a Kondo singlet will form by breaking Cooper pairs at the Fermi level, and then the $0$-junction behavior takes place. Instead, the Josephson junction will enter its $\pi$ phase\cite{jpt4}.
Such a result has been predicted theoretically and observed experimentally, by either the sign change of the supercurrent or the crossing behavior of the Andreev bound states\cite{JoseEXp}. When one QD molecule is introduced to the Josephson junction, the spin correlation, Cooper-pair correlation, and the quantum interference mechanisms will take effect simultaneously, which leads to more interesting Josephson phase transition behaviors. For instance, in a T-shaped double-QD junction, novel transition occurs in the half-filled case, differently from the serially-coupled geometry\cite{TQD,SQD1,SQD2}. In the Fano-Josephson junction, an intermediate bistable phase has been found to appear in the phase-transition process\cite{BSP}.
\par
With respect to the multi-QD structures, QDs can couple to the leads in a parallel way, in addition to forming the QD molecules. One typical case is the well-known parallel double-QD system, in which the Aharonov-Bohm effect can be observed if local magnetic flux is introduced\cite{PQD1,PQD2,PQD3}. When the resonant and nonresonant channels are constructed, the Fano effect has an opportunity to govern the quantum transport result\cite{Fano}. Moreover, in the parallel double-QD system, the strong Coulomb interactions are able to induce the RKKY or $SU(4)$ Kondo effects\cite{RKKY,SU4A,SU4B,SU4C}.
In view of the special properties of the parallel double-QD system, it is natural to think that they can drive interesting Josephson phase transitions. However, such a topic has not been discussed so far. In this Letter, we would like to evaluate the Josephson effect in one parallel double-QD junction with two QDs coupled to the SCs, respectively. After calculation, we see that the uniformity of the two arms weakens the Josephson phase transition, manifested as the fact that the supercurrent only arrives at its $\pi'$ phase with the enhancement of the electron correlation. Such a result reflects the special Josephson phase transition characteristics in the parallel double-QD junction.

\begin{figure}
\begin{center}\scalebox{0.16}{\includegraphics{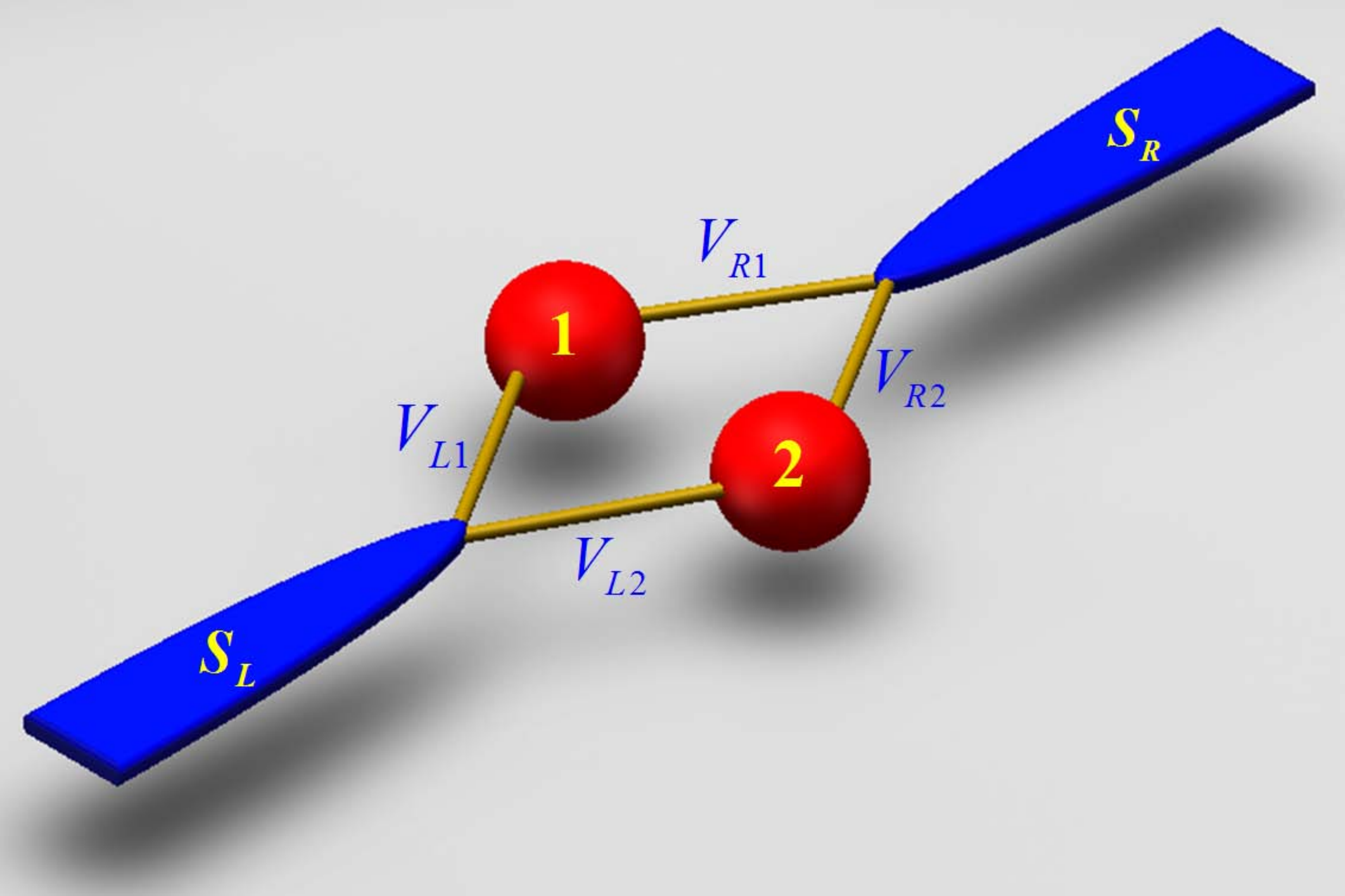}}
\caption{Schematic of a parallel double-QD Josephson junction. The two QDs connect with two $s$-wave SCs, respectively.
\label{Struct}}
\end{center}
\end{figure}
\par
\par
\emph{Theory}---
The Hamiltonian of the parallel Josephson junction is written as $H=H_{S}+H_D+H_{T}$.
The first term is the Hamiltonian of the SCs within the standard BCS mean-field approximation. It takes the form as
\begin{small}
\begin{eqnarray}
H_S&=&\sum_{\alpha k \sigma} \varepsilon_{\alpha k}a^\dag_{\alpha k\sigma}a_{\alpha k\sigma}+\sum_{\alpha k} (\Delta e^{i\varphi_\alpha}a_{\alpha k\downarrow}a_{\alpha -k\uparrow}\notag\\
&&+\Delta e^{-i\varphi_\alpha}a^\dag_{\alpha -k\uparrow}a^\dag_{\alpha k\downarrow}).
\end{eqnarray}
\end{small}
$\varphi_\alpha$ and $\Delta$ are superconducting phase and energy gap, respectively, with $\alpha=L,R$. $a^\dag_{\alpha k\sigma}$($a_{\alpha k\sigma}$) is the operator that creates (annihilates) an electron with energy $\varepsilon_{\alpha k}$ for SC-$\alpha$, where $k$ is the momentum quantum number of the free conduction electrons.
Next, $H_D$, modeling the Hamiltonian for the two QDs, reads
\begin{eqnarray}
H_D&=&\sum_{j\sigma}\varepsilon_{j} d^\dag_{j\sigma}d_{j\sigma}+\sum_{j}U_{j}n_{j\uparrow}n_{j\downarrow}.
\end{eqnarray}
$d^\dag_{j\sigma}$($d_{j\sigma}$) is the operator to create (annihilate) an electron with energy $\varepsilon_j$ and spin $\sigma$ in QD-$j$ ($j=1,2$). $U_j$ indicates the strength of intradot Coulomb repulsion in the corresponding QD. The last term of $H$ denotes the coupling between the QDs and SCs. For our considered system, it can be directly written as
\begin{eqnarray}
H_T=\sum_{j\alpha k\sigma}(V_{\alpha j k}a^\dag_{\alpha k\sigma}d_{j\sigma}+h.c.).
\end{eqnarray}
$V_{\alpha jk}$ describes the QD-SC coupling coefficient.
\par
It is well-known that the phase difference between SCs drives finite current through one Josephson junction. With respect to such a junction, the supercurrent properties can be evaluated by the following
formula $I_J={2e\over\hbar}{\partial \langle H\rangle\over\partial \varphi}={2e\over\hbar}{\partial {\cal F}\over\partial \varphi}$.
$\varphi=\varphi_L-\varphi_R$ is the phase difference between the SCs, and $\langle\cdots\rangle$ is the thermal average. Besides, $\cal F$ is the free energy of the Josephson junction. As one typical case, i.e., zero temperature, $\cal F$ will be simplified to be the ground-state (GS) energy of the system $E_{GS}$. As a result, the supercurrent can be rewritten as
\begin{eqnarray}
I_J={2e\over \hbar}{\partial E_{GS}\over \partial
\varphi}.\label{Jose}
\end{eqnarray}
\par
Note that in such a structure, the GS determination is a formidable task, which usually requires one appropriate approximation scheme, such as the mean-field approximation and zero band-width approximation\cite{ZBWAa,ZBWAb}. However, in comparison with
these methods, the numerical renormalization group (NRG) method is more accurate to reflect the properties of the GS energy\cite{NRG1,NRG2}. We will perform the NRG method to figure out the GS energy. For calculation, we would like to take a few simplifications as follows. The two
SCs are assumed to be identical
($\varepsilon_{Lk}=\varepsilon_{Rk}=\varepsilon_k$ and
$\Delta_L=\Delta_R=\Delta$) except for a finite phase difference
$\varphi=\varphi_L-\varphi_R$; without loss of generality we put
$\varphi_L=-\varphi_R=\varphi/2$. In the normal state, the
conduction bands on the SCs are symmetric with a flat density of
states $\rho_0$ and the bandwidth $\mathscr{D}$ above and below the
Fermi energy. Besides, we only consider the symmetric junction,
$V_{\alpha jk}=V$. The SC-QD coupling is well characterized by the
single parameter $t_0=\pi V^2\rho_0$, which will be fixed $t_0=0.04\mathscr{D}$ throughout this work.

\begin{figure}
\begin{center}\scalebox{0.048}{\includegraphics{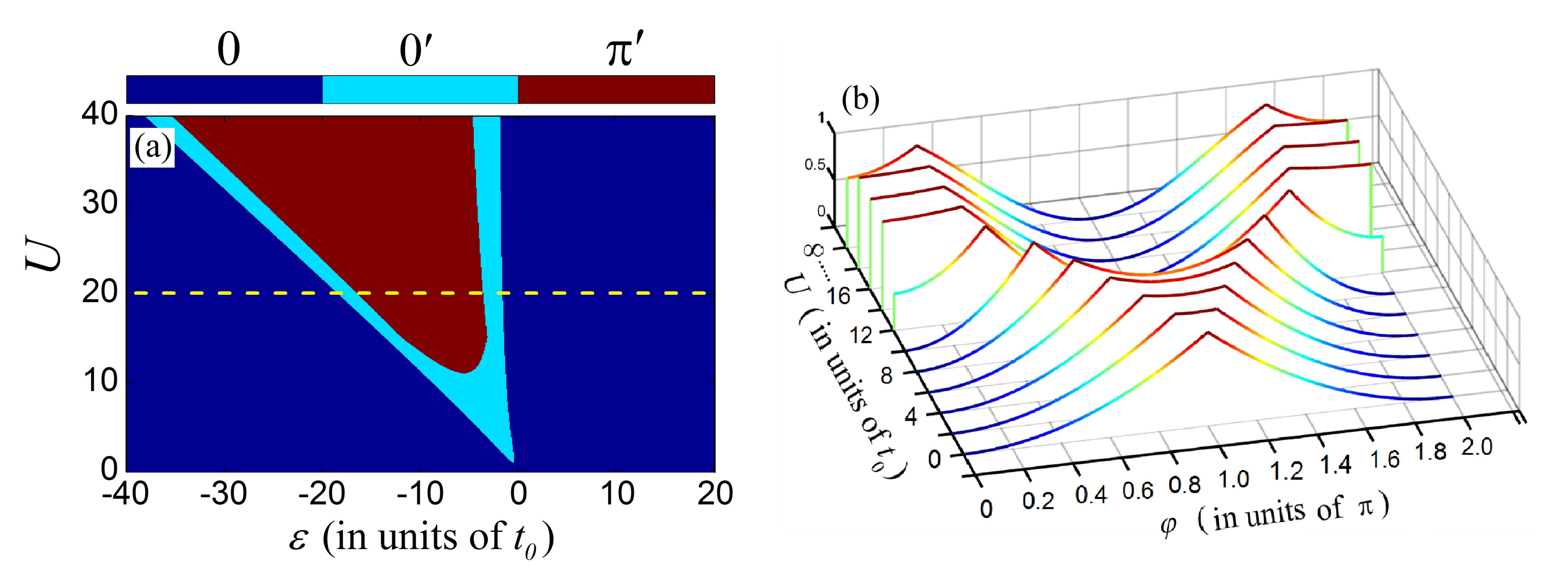}}
\caption{(a) Phase diagram of the Josephson junction with parallel double QDs. (b) GS level of the junction with the increase of intradot Coulomb strength.
\label{Phase1}}
\end{center}
\end{figure}

\par
\emph{Numerical results and discussions}---
With the help of the above theory, we proceed to evaluate the zero-temperature supercurrent in the Josephson junction with parallel double QDs. Before discussion, the phase of the Josephson junction should be defined first. If the system's GS energy as a function of $\varphi$ has a global minimum at the point of $\varphi=0$
($\varphi=\pi$), the junction will be located as its $0$ ($\pi$)
phase. For the $0'$ ($\pi'$) phase, it describes the case where one local minimum emerges at the point of $\varphi=\pi$ ($\varphi=0$) in the $E_{GS}$ spectrum, in addition to the global minimum at $\varphi=0$ ($\varphi=\pi$)\cite{Pai}.

\par
In Fig.2 we take $\varepsilon_j=\varepsilon$ and $U_j=U$ to calculate the supercurrent properties in the parallel double-QD junction. For this purpose, we firstly plot the phase diagram of the supercurrent in Fig.2(a). One can clearly find that such a phase diagram is very similar to the single-arm result, and the phase transition takes place in the region of $-U<\varepsilon<0$ \cite{Pai,singlephase}. However, regardless of the change of the Coulomb interactions or the QD levels, the supercurrent can only arrive at its $\pi'$ phase, but no $\pi$-phase supercurrent comes into being. This means that the parallel double-QD junction exhibits the special  phase-transition behavior. Such a phase transition can be checked by the result in Fig.2(b), which describes the GS energy as a function of the intradot Coulomb interaction in the case of $\varepsilon=-10.0$. It shows that with the increase of $U$, the global minimum of the GS energy moves from the point of $\varphi=0$ to the position of $\varphi=\pi$. Note that in this process, the local minimum of the GS energy always exists the point of $\varphi=0$, and it becomes more apparent with the further increase of Coulomb interaction. Therefore, such a junction can only reach its $\pi'$ phase under the situation of identical QDs and QD-SC couplings.

\begin{figure}[htb]
\begin{center}\scalebox{0.07}{\includegraphics{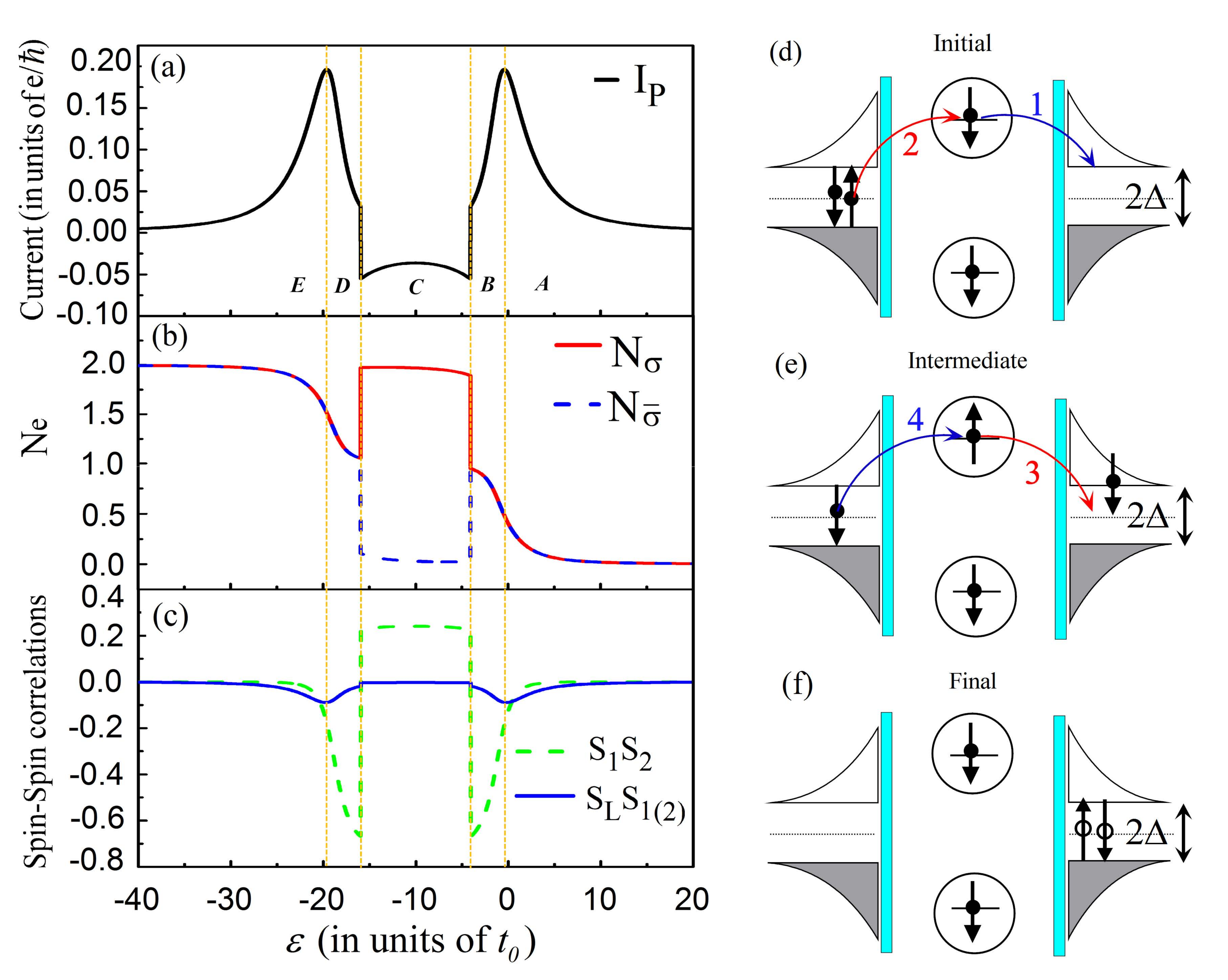}}
\caption{(a)-(c) Spectra of the supercurrent, average electron occupation, and spin correlation with shift of QD levels. Relevant parameters are fixed at $U=20$ and $\varphi={\pi\over2}$. (d)-(f) Illustration of the electron tunneling in region $C$.
\label{tt}}
\end{center}
\end{figure}
\par
In order to further discuss the special Josephson phase transition, we plot the spectra of supercurrent, average electron occupation in the QDs, and spin correlations in this junction. The numerical results correspond to Fig.3(a)-(c), respectively, where relevant parameters are taken to be $\varepsilon_j=\varepsilon$ with $U=20$ and $\varphi={\pi\over2}$. In Fig.3(a), one can find that the supercurrent profile is symmetric about the electron-hole symmetry point $\varepsilon=-{U\over2}$, and two peaks appear at the positions of $\varepsilon=0$ and $\varepsilon=-U$, respectively. For the part of $\varepsilon>-{U\over2}$, it shows that with the decrease of the QD levels to the point of $\varepsilon=0$, the supercurrent magnitude reaches its maximum, whereas the following decrease first suppresses the current magnitude and then reverses the current direction at the position of $\varepsilon\approx-4.0$. Similarly, when the QD levels increase from the region of $\varepsilon<-U$, the same process can also be observed.
\par
The supercurrent spectrum in Fig.3(a) can be clarified with the help of the results in Fig.3(b)-(c). In order to perform discussion, we would like to divide the supercurrent spectrum into five regions according to its variation manner, i.e, regions $\bm A$, $\bm B$, $\bm C$, $\bm D$, and $\bm E$. Firstly, we pay attention to the edge of regions $\bm A$ and $\bm B$ where $\varepsilon=0$. At this point, the total electron number in the QDs is equal to 1.0 with $N_{\sigma}=N_{\bar\sigma}$, and the antiferromagnetic correlation between the QD and SCs reaches its maximum. Thus, it is the QD-SC Kondo correlation that magnifies the $0$-junction behavior. The underlying physics can be understood as follows. When the QD levels (i.e., $\varepsilon_{1(2)}$) are tuned to the energy zero point, both of them will be half occupied and the levels $\varepsilon_j+U$ are empty. In such a case, one can view the parallel double QDs to be one large $\cal QD$ by considering the two arms as the pseudospin indexes. Surely, the uniformity of two arms gives rise to the degeneracy of pseudospin states. One can then find that such a $\cal{QD}$ is singly occupied at the case of $\varepsilon=0$, which drives the occurrence of the Kondo effect. According to the previous works, this is exactly the orbital-Kondo effect. Therefore, the enhanced orbital-Kondo correlation magnifies the Josephson effect. Next, when the QD levels decrease, one new electron is allowed to enter such a large $\cal QD$. The lowest-energy principle certainly demands one spin singlet to form in the $\cal QD$, leading to the weakness of the QD-SC Kondo correlation and the strengthening of the antiferromagnetic correlation between QD-1 and QD-2. As a result, the supercurrent magnitude is suppressed gradually in region $\bf B$.
\par

For the result in region $\textbf{C}$, Fig.3(b) shows that the average electron occupation increases with the decrease of the QD levels, and $N_{\sigma}=2.0$ and $N_{\bar\sigma}=0$. The reason consists in the fact that each QD is occupied by one electron and the two electrons possess identical spin orientation. This certainly leads to the ferromagnetic spin correlation between the QDs. As shown in Fig.3(c), $\langle \textbf{S}_1\cdot\textbf{S}_2\rangle\approx0.3$ in such a region. Since this ferromagnetic correlation arises from the indirect coupling between the two QDs via the SCs, it cannot be viewed as the RKKY correlation. Note, however, that in such a case, the spin-Kondo temperature will be less than the superconducting gap in this region, hence the Cooper-Pair correlation will govern the Josepshon effect. And then, the spin ordering of one Cooper pair can be changed  during its motion process, which induces the $\pi$-junction behavior [See Fig.3(d)-(f)]. It should be noticed that due to the existence of the two identical channels, an electron is allowed to pass through them with the same probability. Accordingly, the nonlocal motion of the electrons in one Cooper pair weakens the $\pi$-junction behavior but only leads to the occurrence of the $\pi'$-phase suppercurrent.
\par
When the QD levels further decrease, levels $\varepsilon_j$ will not provide channels for the Cooper-pair motion, because they are far below the Fermi level of the system. In such a case, the two levels can still be considered to the levels of two opposite-pseudospin states. Due to their degeneracy, the spin singlet should formed. Thus, the strong antiferromagnetic correlation takes place in region $\bm D$. Meanwhile, the levels $\varepsilon_j+U$ begin to get close to the Fermi levels. At the case of $\varepsilon=-U$, such two levels will contribute to the orbital-Kondo effect, leading to the enhancement of the $0$-junction behavior. Next, with the further decrease of the QD levels, both of the QDs will be fully occupied, and then the Josephson effect is weakened in region $\bm E$. Up to now, the supercurrent property in Fig.3(a) has been clarified.

\begin{figure}[htb]
\begin{center}\scalebox{0.048}{\includegraphics{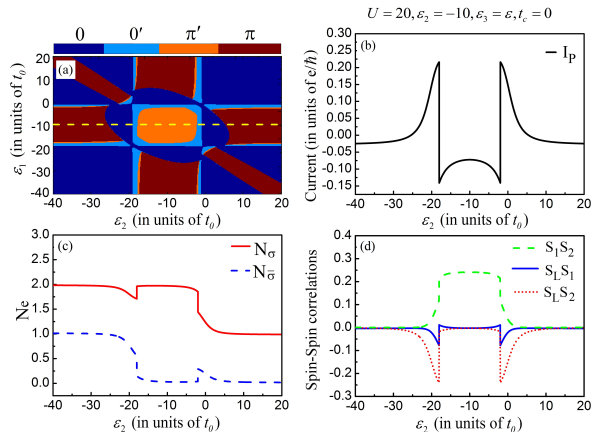}}
\caption{(a) Influence of detuning the QD levels on the phase diagram of the Josephson junction with parallel double QDs. The system's Coulomb strength is taken to be $U=20$. (b)-(d) Spectra of the supercurrent, average electron occupation, and spin-spin correlation in the case of $\varepsilon_2=-{U\over2}$ with $\varphi={\pi\over2}$.
\label{Phasen}}
\end{center}
\end{figure}

\par
In Fig.4, we take $U=20$ and detune the QD levels to further present the supercurrent characteristics. The phase diagram is shown in Fig.4(a). One can clearly observe that this phase diagram exhibits an isolated island of the $\pi'$-phase region, bounded by the lines of $\varepsilon_{1(2)}=0$ and $\varepsilon_{1(2)}=-U$. Outside this island, it is one circular $0$-phase region. Next, with the further detuning of the QD levels, the $\pi$-junction behavior has opportunities to appear under the situations of $\varepsilon_{j}\ll-U$ or $\varepsilon_{j}\gg0$, while $\varepsilon_{j'}$ shifts around the electron-hole symmetry point (i.e., $-U<\varepsilon_{j'}<0$). Surely, such a phenomenon originates from the fact that only one arm contributes to the Cooper-pair tunneling. In recent years, the appearance of the isolated-island region and the $\pi'$ phase
is largely suppressed have become one important interest in the aspect of the Josephson effect modified by the electron correlation mechanism\cite{su4d,SQD2}. One can then ascertain that the result in Fig.4(a) provides new information for understanding the isolated-island behavior of the phase transition.
\par
In addition to the isolated-island behavior, in Fig.4 one can find that abundant phase transition phenomenon takes place when the level of one QD is fixed at the region of $-U<\varepsilon_j<0$. In view of this result, we would like to take $\varepsilon_2=-10$ (i.e., $\varepsilon_2=-{U\over2}$) to reveal the supercurrent property by shifting the level of QD-$1$. The supercurrent curve is shown in Fig.4(b). It shows that when $\varepsilon_2$ decreases from $20$ to $5.0$, the supercurrent direction changes smoothly, hence the direct $\pi\to 0$ phase transition comes into being. Next, in the case of $\varepsilon_2=-2.0$, the positive supercurrent reaches its maximum. Meanwhile, both the direction and magnitude of the supercurrent undergo their sharp change at this position. As a consequence, the Josephson junction enters its $\pi'$ phase. These results can be explained as follows. The decrease of $\varepsilon_2$ causes its-embedded arm to contribute to the Cooper-pair tunneling. To be specific, this leads to the occurrence of the $0$-phase suppercurrent due to the weak electron correlation. The enhancement of such suppercurrent gives rise to the direct $\pi\to 0$ phase transition process. Next, when $\varepsilon_2$ decreases below the system's Fermi level, an electron has a chance to enter QD-2. Accordingly, the QD-SC Kondo correlation will take its effect to the Josephson phase transition [See Fig.4(c)-(d)]. Note that in such a case, the Kondo correlation also exists in the arm with QD-1. The interplay between the two Kondo correlation has an opportunity to induce the RKKY effect, which can be viewed as the reason for the $0'$-phase supercurrent. As the level of QD-2 further decreases, its spin occupation will become robust, similar to QD-1. And then, the $\pi'$-phase behavior comes into play.
\par

\emph{Summary}---
In summary, we have presented an analysis about the Josephson phase transition in a parallel junction in which its each arm has one embedded QD. It has been found that if the two QDs are half occupied, the Josephson phase transition will become weak. To be concrete, with the enhancement of the electron correlation, the supercurrent can only arrive at its $\pi'$ phase but does not enter its $\pi$ phase. Via analysis, we consider such a result to be caused by the nonlocal motion of the two electrons of one Cooper pair in its motion process. In addition, the isolated-island behavior has been observed when the QD levels are detuned. Based on the obtained results, we believe that this work can be helpful in understanding the Josephson phase transition modified by interaction between the electron correlation and the quantum interference.

\section*{Acknowledgments}
\par
This work was financially supported by the Fundamental Research
Funds for the Central Universities (Grant No. N160504009) and the National Natural Science Foundation of China (Grant No. 11604221). Our numerical results are obtained via ``NRG Ljubljana"---open source numerical renormalization group code.

\clearpage

\bigskip

\end{document}